\documentclass[twocolumn]{aastex631}

\usepackage{gensymb}
\usepackage{textcomp}
\usepackage{mathtools}
\usepackage{booktabs}
\usepackage{amsmath}	
\usepackage{comment}

\accepted{March 27, 2026}
\submitjournal{ApJL}

\shorttitle{Loktak: LAE Protocluster at $z \simeq 4.9$}
\shortauthors{Laishram et al.}

\begin{document}

\title{Discovery of a $z\simeq 4.9$ Lyman-$\alpha$ Emitter Protocluster: Wavelength-Dependent Environmental Effects on Galaxy Structure}  

\correspondingauthor{Ronaldo Laishram}
\email{ronaldo.laishram@nao.ac.jp}

\author[0000-0002-0322-6131]{Ronaldo Laishram}
\affiliation{National Astronomical Observatory of Japan,
2-21-1 Osawa, Mitaka, Tokyo 181-8588, Japan}

\author[0000-0002-0479-3699]{Yusei Koyama}
\affiliation{National Astronomical Observatory of Japan,
2-21-1 Osawa, Mitaka, Tokyo 181-8588, Japan}
\affiliation{Department of Astronomical Science,
The Graduate University for Advanced Studies,
2-21-1 Osawa, Mitaka,
Tokyo 181-8588, Japan}

\author[0000-0002-3801-434X]{Haruka Kusakabe}
\affiliation{Department of General Systems Studies, Graduate School of Arts and Sciences, The University of Tokyo, 3-8-1 Komaba, Meguro-ku, Tokyo, 153-8902, Japan}

\author[0000-0003-3214-9128]{Satoshi Kikuta}
\affiliation{Department of Astronomy, School of Science, The University of Tokyo, 7-3-1, Hongo, Bunkyo, Tokyo 113-0033, Japan}

\author[0009-0001-9612-1223]{Shunta Shimizu}
\affiliation{Department of Astronomy, School of Science, The University of Tokyo, 7-3-1, Hongo, Bunkyo, Tokyo 113-0033, Japan}

\author[0000-0002-2993-1576]{Tadayuki Kodama}
\affiliation{Astronomical Institute, Tohoku University, 6-3, Aramaki, Aoba, Sendai, Miyagi 980-8578, Japan}

\begin{abstract}
We report the discovery of a Lyman-$\alpha$ emitter (LAE) protocluster at $z = 4.90$ in the COSMOS field, comprising four distinct overdensity peaks spanning $\sim$65 $\times$ 36 cMpc$^2$, with the primary concentration exhibiting a 4-fold surface density enhancement relative to the field within a 1.5 proper Mpc (pMpc) radius. Using SILVERRUSH narrowband survey data combined with JWST COSMOS-Web imaging, we perform one of the first systematic rest-frame optical and UV morphological comparisons of protocluster versus field LAEs at this redshift using JWST NIRCam rest-frame UV (F150W, $\sim$2540\AA) and optical (F277W, $\sim$4700\AA) imaging. S\'ersic profile fitting for 16 protocluster members and 23 field LAEs reveals a size difference: protocluster LAEs are $\sim$40\% larger in rest-optical (median $R_e = 0.81_{-0.04}^{+0.26}$ kpc vs.\ $0.58_{-0.04}^{+0.11}$ kpc, $p = 0.041$) with no significant difference in rest-UV ($p = 0.51$) or S\'ersic index. At fixed stellar mass, protocluster LAEs are offset by $+0.12$~dex ($\simeq$31\%) in rest-optical size from the field size-mass relation (68\% CI: $[+0.08, +0.21]$; Mann-Whitney $p = 0.033$), with 75\% exhibiting positive size residuals compared to 44\% of field LAEs. This wavelength-dependent environmental signature suggests that protocluster environments at $z \simeq 5$ preferentially affect extended stellar populations, possibly through tidal interactions, with no significant environmental difference detected in rest-UV sizes, providing observational evidence for environmental influences on the structure of LAEs during the early build-up phase of cosmic star formation.
\end{abstract}


\section{\textbf{Introduction}}
\label{sec:intro}

Galaxy protoclusters represent the progenitors of the most massive structures in the present-day Universe, providing critical laboratories for investigating the onset of environmental influences on galaxy evolution. While the morphology-density relation is well-established in local clusters \citep{Dressler1980}, determining when environmental processes begin to shape galaxy structure remains a fundamental question in observational cosmology. At $z \simeq 5$, during the early build-up phase of cosmic star formation, protoclusters occupy a unique position where hierarchical assembly is ongoing and are predicted to evolve into massive clusters by $z=0$ \citep{Chiang_et_al_2017, Ata_et_al_2022}. In these dense regions, mechanisms such as enhanced mergers, tidal interactions, and gas accretion are predicted to leave their first observable signatures on member galaxies \citep{Boselli_and_Gavazzi_2006, Overzier_2016}. Identifying and characterizing these structures at such early epochs is therefore essential for constraining when dense environments begin to regulate galaxy properties.

Lyman-$\alpha$ emitters (LAEs) serve as highly efficient tracers of protoclusters due to their ubiquity and established clustering properties at early cosmic times \citep{Ouchi_et_al_2018, Kusakabe_et_al_2018, Higuchi_et_al_2019, Ouchi_et_al_2020, Hu_et_al_2021, Ramakrishnan_et_al_2024}.  Narrowband imaging surveys have successfully mapped LAE overdensities across a wide range of epochs \citep{Lee_et_al_2014,  Badescu_et_al_2017, Lee_et_al_2024}. \citet{Shimasaku_et_al_2003} discovered the first proto-large-scale structure traced by LAEs at $z \approx 4.86$ in the Subaru Deep Field, revealing an elongated overdense region spanning $\sim$65 comoving Mpc. Subsequent surveys have extended these discoveries across a wide redshift range, identifying protoclusters at $z = 2.3$ \citep{Badescu_et_al_2017}, $z = 2.47$ \citep{Huang_et_al_2022}, $z = 5.3$ \citep{Capak_et_al_2011}, and $z \simeq 4.5$ \citep{Staab_et_al_2024, Rubet_et_al_2025}. At the highest redshifts approaching the epoch of reionization, \citet{Hu_et_al_2021} identified LAGER-z7OD1 at $z \approx 6.9$, while \citet{Ramakrishnan_et_al_2023} demonstrated that LAEs and Lyman-$\alpha$ blobs preferentially reside in overdense regions associated with protoclusters. These structures exhibit characteristic surface density enhancements that are approximately 3--6 times higher than those in field regions.

While the spatial clustering of LAEs has been extensively characterized, their morphological properties, particularly in dense protocluster environments, remain poorly constrained at $z \gtrsim 4$. Previous \textit{HST} studies have revealed that LAEs are compact systems with effective radii ranging from $\sim$100--200~pc at $z \simeq 2$--3 \citep{Kim_et_al_2026} to $R_e \lesssim 1$~kpc at higher redshifts \citep{Bond_et_al_2012, Shibuya_et_al_2019}. While studies of the Ly$\alpha$ luminosity function and equivalent width distribution in protoclusters at $z \simeq 3.8$ find these properties to be largely independent of environment \citep{Malavasi_et_al_2021}, structural properties may be more sensitive to the frequent mergers and interactions characterizing these dense regions. Furthermore, \textit{HST} observations at $z \gtrsim 5$ are restricted to rest-frame UV wavelengths, which trace recent star formation but may not reflect the underlying stellar mass distribution shaped by environmental processes. Although numerous studies have identified LAE overdensities at $z \simeq 4$--6, quantitative comparisons of rest-frame optical structural parameters (e.g., effective radii, S\'ersic indices) between protocluster and field LAEs remain absent at these redshifts.

The advent of JWST has transformed this landscape by enabling high-resolution imaging at rest-frame optical wavelengths for $z > 4$ galaxies \citep{Casey_et_al_2023}, allowing direct measurements of stellar mass distributions in high-redshift protoclusters and decoupling young UV-bright star-forming regions from extended older stellar populations. Recent studies suggest that overdensities may accelerate the emergence of evolved galaxy properties even at these early epochs \citep{Morishita_et_al_2025}. Determining whether galaxy sizes and structural parameters depend on local density at these early epochs is critical for establishing when and how environmental mechanisms begin to regulate stellar mass assembly.

In this Letter, we report the discovery of an LAE protocluster at $z = 4.90$ in the COSMOS field, identified from the Systematic Identification of LAEs for Visible Exploration and Reionization Research Using Subaru HSC (SILVERRUSH) narrowband survey \citep{Kikuta_et_al_2023}. The structure comprises four distinct overdensity peaks spanning $\sim$65 $\times$ 36 cMpc$^2$, with surface density enhancements of up to a factor of 4 relative to the field within 1.5~pMpc radii of each peak. We refer to this system as the Loktak protocluster.\footnote{The name is inspired by Loktak Lake in Manipur, India, whose floating islands within an interconnected body of water evoke the multi-component nature of this large-scale structure.} Leveraging deep NIRCam imaging from the JWST COSMOS-Web treasury survey \citep{Casey_et_al_2023, Franco_et_al_2024}, we perform one of the first systematic rest-frame optical structural comparisons of protocluster versus field LAEs at $z \simeq 5$. We measure effective radii in rest-frame UV (F150W) and rest-frame optical (F277W) through S\'ersic profile fitting to investigate wavelength-dependent structural properties. Throughout this work, we adopt a flat $\Lambda$CDM cosmology with $\Omega_m = 0.3$, $\Omega_\Lambda = 0.7$, and $H_0 = 70$~km~s$^{-1}$~Mpc$^{-1}$, and quote magnitudes in the AB system.

\section{\textbf{Data}} \label{sec:data}

\subsection{Ly$\alpha$ Emitter Sample} \label{sec:lae_sample}

Our parent sample of LAEs at $z \simeq 4.9$ is drawn from the SILVERRUSH program \citep{Ouchi_et_al_2018, Shibuya_et_al_2018, Kikuta_et_al_2023}, which identified LAEs using narrowband (NB) imaging from the Hyper Suprime-Cam Subaru Strategic Program (HSC-SSP; \citealt{Aihara_et_al_2022}) and the Cosmic HydrOgen Reionization Unveiled with Subaru (CHORUS) survey \citep{Inoue_et_al_2020}. The HSC-SSP Deep layer covers an effective area of $\sim$25~deg$^2$ using HSC on the 8.2~m Subaru Telescope \citep{Aihara_et_al_2019, Aihara_et_al_2022}. CHORUS complements HSC-SSP with deeper NB imaging in the COSMOS field using custom filters including NB718. The combined SILVERRUSH XIII catalog \citep{Kikuta_et_al_2023} contains 726 photometrically selected LAEs in the NB718 filter ($\lambda_c = 7170$~\AA, FWHM $= 111$~\AA) covering 1.76~deg$^2$ in the COSMOS field, targeting Ly$\alpha$ emission at $z = 4.90 \pm 0.046$.

LAE candidates were selected via narrowband excess (NB $-$ BB color) with $>5\sigma$ detection significance in NB718, combined with Lyman break color criteria using multi-band photometry to reject low-redshift interlopers \citep{Ouchi_et_al_2018, Kikuta_et_al_2023}. Source detection and photometry were performed on PSF-matched coadded images processed with hscPipe version 8. Full details are provided in \citet{Kikuta_et_al_2023}. The SILVERRUSH catalog includes 13 spectroscopically confirmed LAEs at $z \simeq 4.9$ in the COSMOS field, validating the photometric selection.

\subsection{JWST COSMOS-Web Imaging Data} \label{sec:jwst_data}

We utilized deep near-infrared imaging from the COSMOS-Web program \citep{Casey_et_al_2023}, a JWST Cycle 1 treasury survey that mapped 0.54~deg$^2$ of the COSMOS field with NIRCam. The NIRCam observations employ four broadband filters---F115W, F150W, F277W, and F444W---reaching $5\sigma$ point-source sensitivities of 26.6--27.3~mag, 26.9--27.7~mag, 27.5--28.2~mag, and 27.5--28.2~mag, respectively, measured in 0\farcs15 diameter apertures.

Observations were conducted between 2023 January and 2024 May. Full details of the image processing and mosaic construction are provided by \citet{Franco_et_al_2024, Franco_et_al_2025}. The final NIRCam mosaics are available at pixel scales of both 30~mas and 60~mas; we adopt the higher-resolution 30~mas mosaics for our structural measurements to optimize spatial sampling of compact high-redshift sources.

At $z = 4.90$, the NIRCam filters probe rest-frame wavelengths of approximately 1950~\AA\ (F115W), 2540~\AA\ (F150W), and 4700~\AA\ (F277W). The F150W filter traces rest-frame UV continuum emission dominated by young stellar populations, while F277W samples rest-frame optical light sensitive to stellar mass and older populations.

\subsection{Photometric Catalog and Derived Physical Properties} \label{sec:phot_catalog}

We obtained stellar masses and star formation rates (SFRs) from the COSMOS-Web photometric catalog presented by \citet{Shuntov_et_al_2025}, which provides multi-wavelength photometry and spectral energy distribution (SED) fitting results for all detected sources in the survey footprint. The catalog employs the LEPHARE SED-fitting code \citep{Arnouts_et_al_2002, Ilbert_et_al_2006} to derive photometric redshifts and physical parameters from composite photometry spanning 32 bands from $\sim$0.3 $\mu$m to $\sim$8 $\mu$m \citep{Shuntov_et_al_2025}.

Template fitting is based on \citet{Bruzual_Charlot_2003} stellar population synthesis models incorporating varied star formation histories, stellar ages, metallicities, and dust attenuation prescriptions \citep{Calzetti_et_al_2000}. For each source, LEPHARE computes a probability density function of the photometric redshift; physical properties (stellar mass, SFR) are then calculated at the median photometric redshift. \citet{Shuntov_et_al_2025} demonstrate that LEPHARE-derived stellar masses show good agreement with independent CIGALE estimates \citep{Boquien_et_al_2019}, providing confidence in the adopted values. We adopt stellar masses from the LEPHARE fits for consistency with the published COSMOS-Web catalog.

\subsection{Spectroscopic Redshift Compilation} \label{sec:spec_catalog}

To validate photometric redshifts and identify potential contamination in our LAE sample, we cross-referenced our sources with the COSMOS spectroscopic redshift compilation of \citet{Khostovan_et_al_2026}, which aggregates measurements from 138 spectroscopic programs spanning the COSMOS field from the local Universe to $z \simeq 8$.

We cross-matched our NB718 LAE catalog with the COSMOS-Web NIRCam imaging using a matching radius of 0\farcs5 following \citet{Shuntov_et_al_2025}, yielding 215 initial LAE candidates with JWST detections. To ensure sample purity, we implemented a two-stage cleaning procedure. First, we cross-referenced all candidates with the COSMOS spectroscopic redshift compilation \citep{Khostovan_et_al_2026} to identify and remove contaminants confirmed at discrepant redshifts. Within the 0.54 deg$^2$ COSMOS-Web survey footprint, this compilation contains 49 spectroscopically confirmed galaxies at $z = 4.90 \pm 0.046$ (the NB718 transmission window) with quality flags $\geq 3$ (reliable or very secure redshifts); of these, eight are positionally matched to our LAE candidates, providing direct validation of the narrowband selection. Second, we applied photometric redshift quality cuts using the COSMOS-Web LEPHARE-derived redshifts \citep{Shuntov_et_al_2025}, retaining only sources with photometric redshifts consistent with the NB718 selection window within $\Delta z \leq 0.2$. This threshold balances sample completeness against contamination from photometric redshift outliers, and we confirmed that our results are insensitive to moderate variations in this criterion. After applying these cuts, our final clean sample comprises 177 LAEs with robust redshift confirmation. All eight spectroscopically confirmed LAEs pass our photometric redshift quality cuts and are retained in the final clean sample.

%==============================================================================
\section{\textbf{Large-Scale Overdensity Structure at $z = 4.90$}} \label{sec:overdensity}

To characterize the environment of LAEs with JWST coverage within the 0.54 deg$^2$ COSMOS-Web survey area, we measured local galaxy overdensity using two independent methods following established protocols \citep{Ramella_et_al_2001, Cooper_et_al_2005, Darvish_et_al_2015}: (1) the fifth nearest neighbor distance, which measures the projected distance to the fifth nearest LAE and calculates local surface density as $\sigma_5 = 5/(\pi d_5^2)$ [Mpc$^{-2}$], normalized to the dimensionless overdensity $\delta_{\rm NN} = \sigma_5 / \langle\sigma_5\rangle - 1$; and (2) Voronoi tessellation, where each LAE defines a cell representing the region closer to it than to any other galaxy, with local density $\rho_{\rm V} = 1/A_{\rm cell}$ [Mpc$^{-2}$] normalized as $\delta_{\rm V} = \rho_{\rm V} / \langle\rho_{\rm V}\rangle - 1$.  Edge effects in the overdensity measurements are mitigated because the COSMOS-Web footprint is embedded within the larger SILVERRUSH/HSC survey area, ensuring all LAEs have complete neighbor coverage for environment calculations. Both methods yield consistent overdensity measurements across the field.

\begin{figure*}

\includegraphics[width=0.99\textwidth]{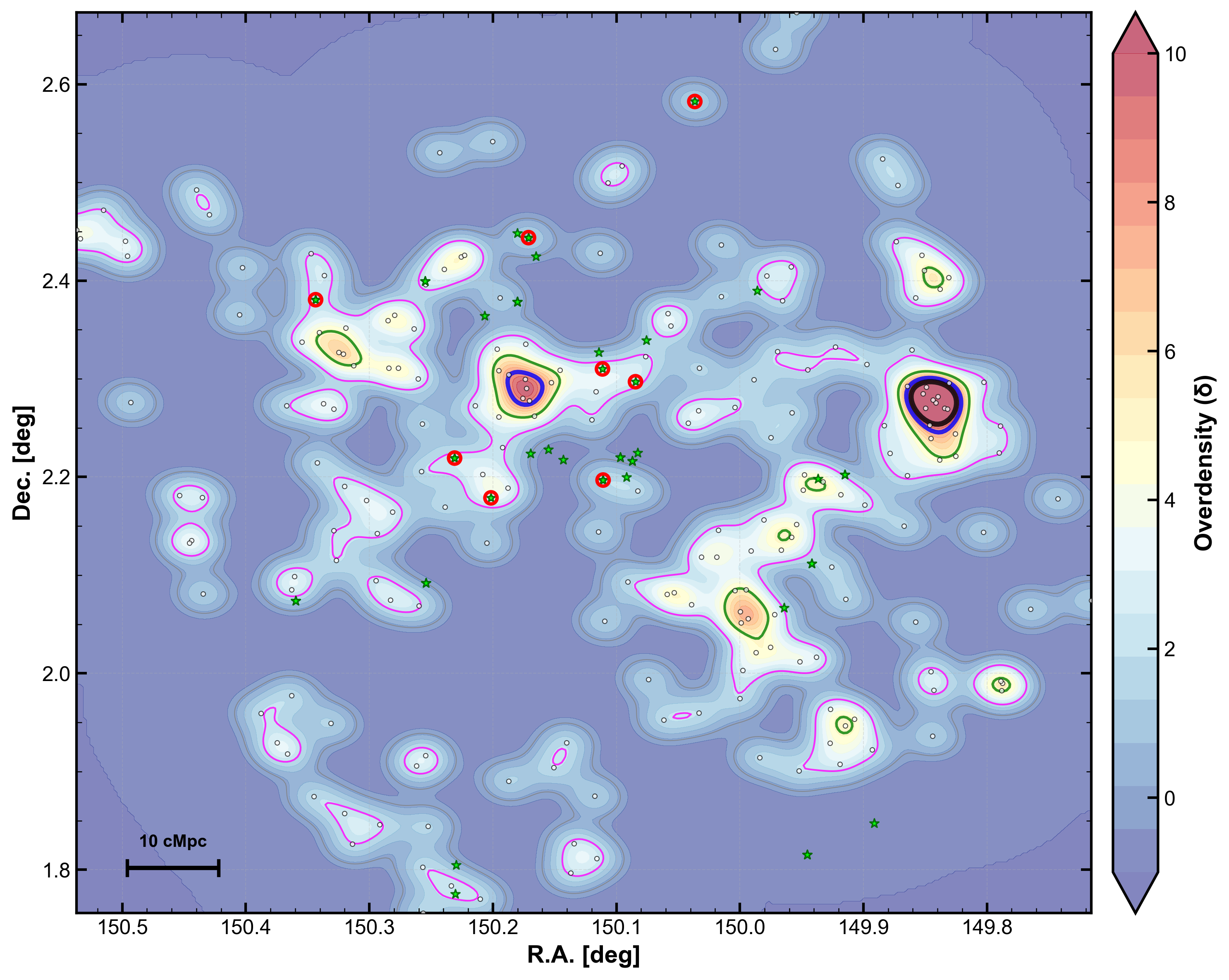}

\caption{Overdensity map of NB718 LAEs at $z = 4.90$ revealing a large-scale protocluster structure (the Loktak protocluster) within the 0.54 deg$^2$ COSMOS-Web survey area. The background shows the KDE-smoothed density field (colorbar: $\delta$), with warmer colors indicating higher overdensity. Contour lines trace overdensity levels at $\delta = 2$ (magenta), $\delta = 5$ (green), $\delta = 8$ (blue), and $\delta = 10$ (black). Individual LAE positions are shown as white open circles. Spectroscopically confirmed galaxies at $z \simeq 4.9$ are marked as green stars, while those cross-matched with our LAE sample (within 0.5\arcsec) are further highlighted with red open circles overlaid on the stars. The structure exhibits a prominent primary peak (4-fold surface density enhancement within 1.5 pMpc radius), along with three additional significant peaks at distances of 7.6, 6.0, and 11.1~pMpc from the center, forming an interconnected protocluster complex spanning $\sim$65 $\times$ 36 cMpc$^2$ characteristic of large-scale structure at this epoch.}

\label{fig:overdensity_map}

\end{figure*}

The resulting spatial distribution reveals a prominent large-scale overdensity structure at $z \simeq 4.9$ (the Loktak protocluster). Figure~\ref{fig:overdensity_map} presents the overdensity map constructed from these 177 LAEs. The structure exhibits a clear concentration of LAEs with multiple interconnected components, characteristic of protocluster environments in the early Universe.

To visualize the full spatial structure and identify overdensity peaks, we constructed a smoothed density field using Gaussian Kernel Density Estimation (KDE) following methods for mapping high-redshift structures \citep{LeeKS_et_al_2014, Badescu_et_al_2017, Huang_et_al_2022}. We applied KDE with a bandwidth scaling factor of 0.3 to balance structure detail against noise, evaluated on a 300$\times$300 grid, and smoothed with a 2D Gaussian filter ($\sigma = 1.5$ pixels). The primary peak, corresponding to the highest density concentration, is located at RA $= 149.8412^\circ$, Dec $= 2.2755^\circ$.

The smoothed density map reveals four distinct peaks spanning $\sim$65 $\times$ 36 cMpc$^2$ on the sky. The primary peak exhibits the highest concentration. Three additional significant peaks are located at 7.6~pMpc to the east, 6.0~pMpc to the southeast, and 11.1~pMpc to the east, forming an extended structure suggestive of filamentary connections between density concentrations. Within 1.5~pMpc of the primary peak, 18 LAEs are enclosed, yielding a surface density of $(36.9 \pm 8.7) \times 10^{-2}$ arcmin$^{-2}$, signifying a 4-fold increase compared to the overall surveyed region ($(9.11 \pm 0.62) \times 10^{-2}$ arcmin$^{-2}$ for 177 LAEs over 0.54 deg$^2$); this enhancement rises to 5.5 times within 1.0~pMpc, indicating a steeply rising density profile toward the center. The three additional peaks show similar enhancements of 3, 2.8, and 2.5 times within their respective 1.5~pMpc radii, confirming the multi-component nature of the protocluster complex in an active phase of hierarchical assembly. Isodensity contour analysis independently confirms all four peaks with high statistical significance (4--12$\sigma$), demonstrating the robustness of the structure identification. As a robustness check, we verified the overdensity significance using the full SILVERRUSH NB718 catalog (726 photometrically selected LAEs over 1.76~deg$^2$), without restriction to the COSMOS-Web footprint. The primary peak remains a highly significant overdensity ($6.5\sigma$ Poisson significance), consistent with the $\sim$4-fold enhancement derived from the photometrically cleaned COSMOS-Web sample. The primary peak also remains the highest-density region in the KDE map constructed from the full-field catalog, confirming that the detection of the Loktak protocluster is not sensitive to the restriction of the overdensity analysis to the COSMOS-Web area.

We estimate the present-day descendant mass $M_{z=0}$ of each overdensity peak following the bias-corrected cylinder method of \citet{Chiang_et_al_2013} and \citet{Hu_et_al_2021}.
For each peak we count LAEs within a cylinder of transverse radius $r = 1.5$~pMpc and line-of-sight depth $\Delta d_\mathrm{LOS} = 49.5$~cMpc (set by the NB718 filter FWHM $= 111$~\AA\ at $z = 4.90$; \citealt{Kikuta_et_al_2023}), giving a comoving volume of $V_\mathrm{cyl} = 1.22 \times 10^4$~cMpc$^3$.
We use the full SILVERRUSH NB718 catalog (726 LAEs, 1.76~deg$^2$) for both the peak counts and the mean density $\bar{n} = 4.59 \times 10^{-4}$~cMpc$^{-3}$, computed from the full survey footprint to avoid biasing $\bar{n}$ upward by the protocluster itself.
The galaxy overdensity $\delta_g = (N_i/V_\mathrm{cyl} - \bar{n})/\bar{n}$ is converted to matter overdensity $\delta_m$ via $1 + b\delta_m = C(1 + \delta_g)$, where $C = 1 + f - f(1+\delta_m)^{1/3}$ corrects for redshift-space distortions \citep{Peebles1980} and $f = \Omega_m(z)^{4/7} = 0.994$ at $z = 4.90$.
We adopt galaxy bias $b = 3.7$ from \citet{Kovac_et_al_2007}, consistent with $b = 4.1 \pm 0.2$ at $z = 5.7$ \citep{Ouchi_et_al_2018} and independently confirmed by $b = 3.69 \pm 0.40$ at $z = 4.9$ from angular clustering of the same NB718 catalog \citep{Umeda_et_al_2025}.
The present-day mass of each peak is $M_{z=0} = (1 + \delta_m)\,\bar{\rho}_m\,V_\mathrm{cyl}$, where $\bar{\rho}_m = 4.08 \times 10^{10}\,M_\odot\,\mathrm{Mpc}^{-3}$ (comoving mean matter density).
The four peaks yield per-peak descendant masses of $7.85$, $7.01$, $6.65$, and $6.65 \times 10^{14}\,M_\odot$ at our adopted bias $b = 3.7$. The dominant systematic uncertainty is the assumed galaxy bias $b$. Varying $b$ over the plausible range $1.5$--$7.0$, the per-peak masses span $(0.6$--$1.1) \times 10^{15}\,M_\odot$, placing each peak in the mass range typically associated with galaxy cluster progenitors.

Based on the overdensity measurements, we classify galaxies into distinct environmental samples for subsequent analysis. We define protocluster members as LAEs within a projected radius of 3.0~pMpc from the primary peak of the smoothed density field, yielding 28 candidate member LAEs. We confirmed that our results are consistent when adopting smaller membership radii (1.5 and 2.0~pMpc), though the larger radius provides a more statistically robust sample. We additionally verified the result using larger membership radii (4.0 and 5.0~pMpc); details are given in Section~\ref{sec:results_morph}. Field LAEs are selected from the lowest density regions, defined as the bottom 25th percentile of the nearest neighbor overdensity distribution ($\delta_{\rm NN}$), yielding 44 galaxies. We focus on the primary peak as it exhibits the strongest overdensity ($\sim$4 times within 1.5~pMpc; Section~\ref{sec:overdensity}); hereafter, we refer to these LAEs as protocluster members and the low-density LAEs as field LAEs for all subsequent analysis.

To investigate environmental effects on galaxy structure, we measured morphologies of LAEs in the protocluster and field environments using {\tt GALFIT} \citep{Peng_et_al_2002}. We fit single S\'ersic profiles to JWST NIRCam imaging in two filters: F150W, sampling rest-frame UV ($\sim$2540\AA), and F277W, sampling rest-frame optical ($\sim$4700\AA) at $z = 4.90$. Free parameters included the effective radius ($R_e$, containing half the total light), S\'ersic index ($n$, describing profile concentration), total magnitude, and centroid position. Each fit was convolved with an empirical PSF constructed from stacking unsaturated stars in the field. We retain only galaxies whose fitted half-light radii exceed the PSF half-width at half-maximum (HWHM), ensuring the objects are sufficiently resolved for reliable S\'ersic measurements. After applying this resolution cut, we retained 17 protocluster and 32 field LAEs in F150W, and 16 protocluster and 23 field LAEs in F277W with reliable size measurements. These final samples result from two sequential selections (a projected distance criterion and a PSF resolution cut), which limits the statistical power of the comparison. We tested the sensitivity of the results to both steps by varying the membership radius and by performing a sensitivity test that includes unresolved sources assigned the PSF resolution limit as an upper bound on their size; details are given in Section~\ref{sec:results_morph}. We verified our size measurements using independent fitting with {\tt Galight} \citep{Ding_et_al_2020} and {\tt PetroFit} \citep{Geda_et_al_2022}, confirming consistent size estimates across methodologies (detailed morphological analysis in Laishram et al., in preparation). Effective radii were converted from angular to physical units (kpc) using the angular diameter distance at $z = 4.90$. Uncertainties on median values are estimated via non-parametric bootstrap resampling (10,000 iterations; 68\% confidence interval from the 16th--84th percentiles of the bootstrap distribution); for two-sample comparisons, both samples are resampled jointly in each iteration to derive uncertainties on the difference of medians. Statistical significance is assessed using two-sided Mann-Whitney $U$ and Kolmogorov-Smirnov tests.

\begin{figure*}
\includegraphics[width=0.49\textwidth]{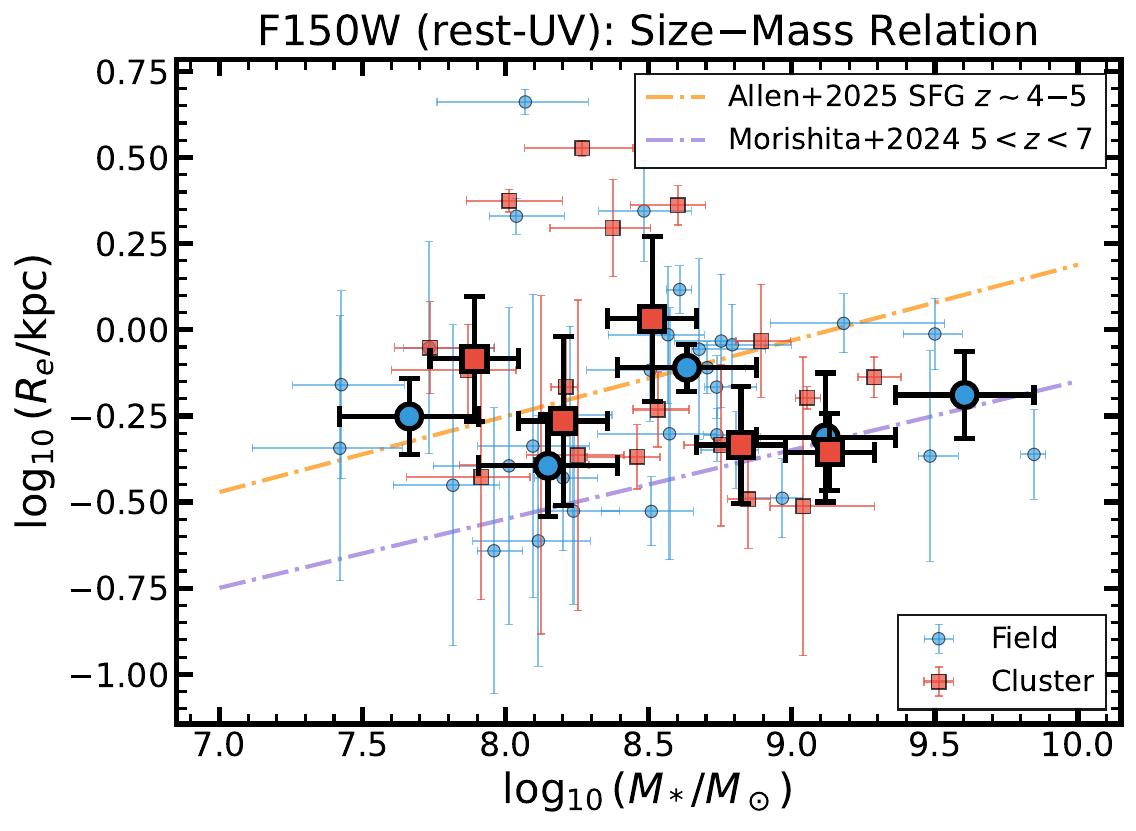}
\includegraphics[width=0.49\textwidth]{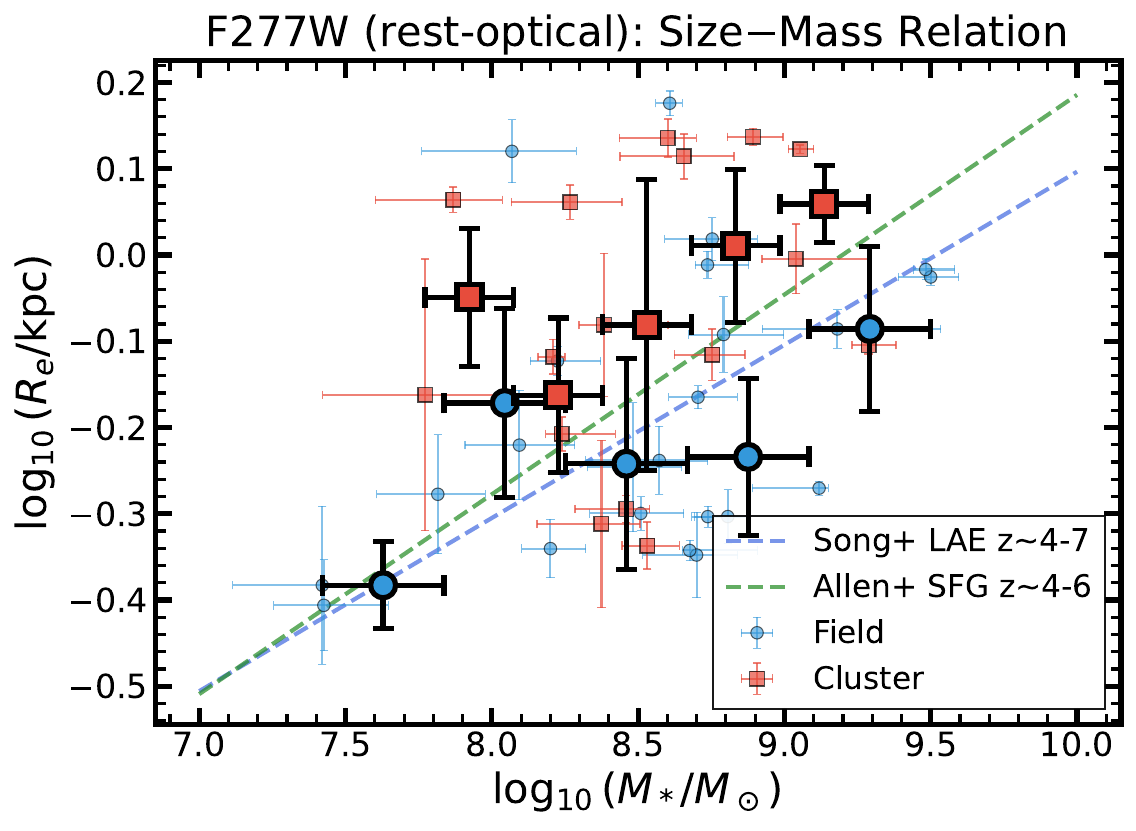}

\caption{Size-mass relations for LAEs in rest-UV (F150W, left panel) and rest-optical (F277W, right panel). Blue circles represent field LAEs, and red squares represent protocluster members. Large filled symbols show binned medians with bootstrap uncertainties. In the left panel (rest-UV), no systematic offset is detected between protocluster and field LAEs (field fit: $\beta = 0.05 \pm 0.08$, scatter $= 0.2$~dex). For reference, the left panel includes rest-UV size-mass relations from \citet{Allen_et_al_2025} for star-forming galaxies (SFGs) at $4 \leq z < 5$ (orange dash-dot line; from their Table~A.1, F150W) and from \citet{Morishita_et_al_2024} for galaxies at $5 < z < 14$ (purple dash-dot line; evaluated at $z = 4.9$). In the right panel (rest-optical), protocluster LAEs systematically lie above the field relation, indicating larger sizes at fixed stellar mass. The median rest-optical size enhancement is $\simeq$40\% (cluster median $R_e = 0.81$~kpc vs.\ field $0.58$~kpc). For reference, the right panel includes the rest-optical size-mass relations from \citet{Song_et_al_2026} for LAEs at $z \simeq 4\mbox{--}7$ (blue dashed line) and from \citet{Allen_et_al_2025} for SFGs at $z \simeq 4\mbox{--}6$ (green dashed line). Field LAEs are broadly consistent with the \citet{Song_et_al_2026} LAE relation (Section~\ref{sec:results_sizemass}).}

\label{fig:size_mass}

\end{figure*}

\begin{figure*}
\includegraphics[width=0.49\textwidth]{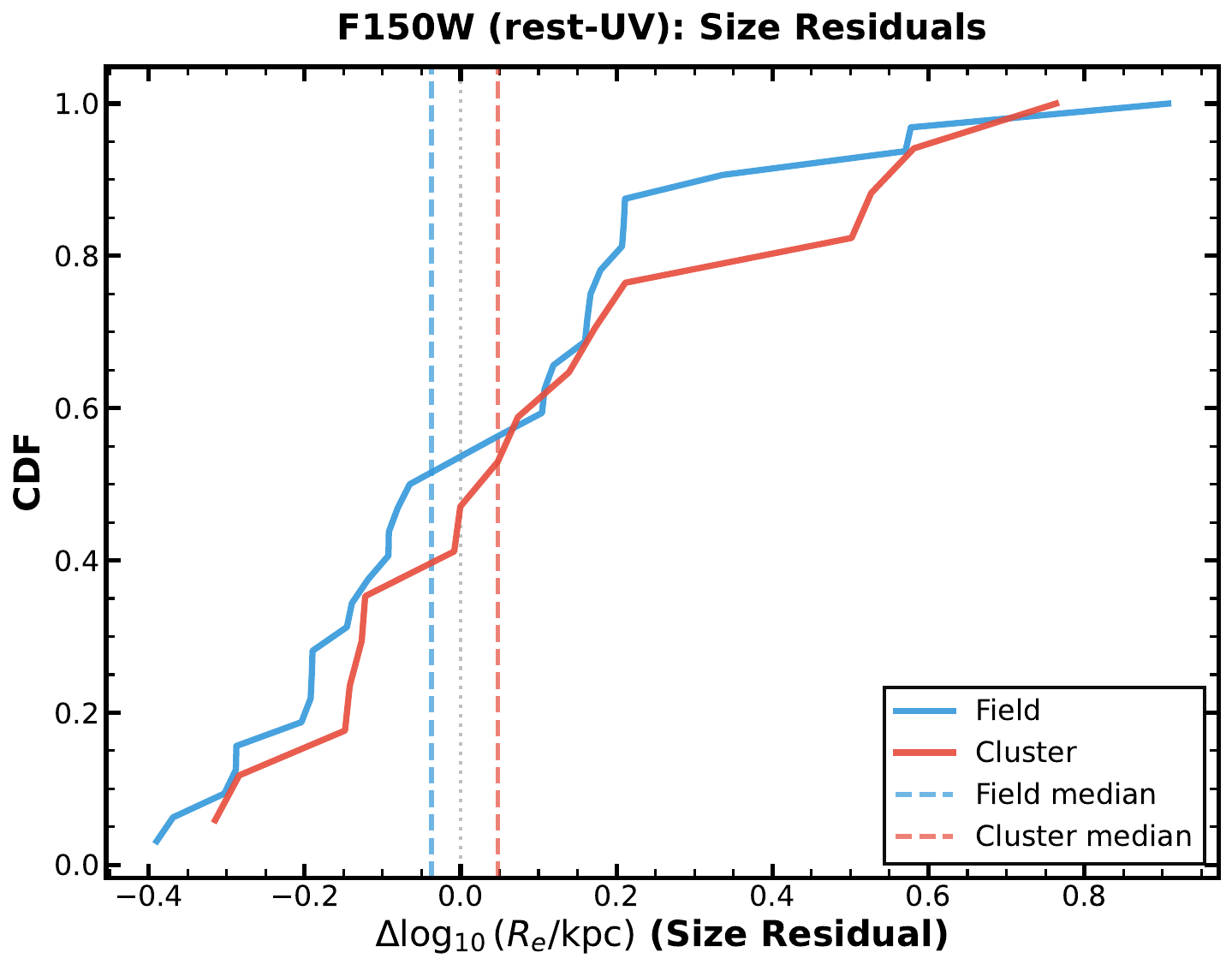}
\includegraphics[width=0.49\textwidth]{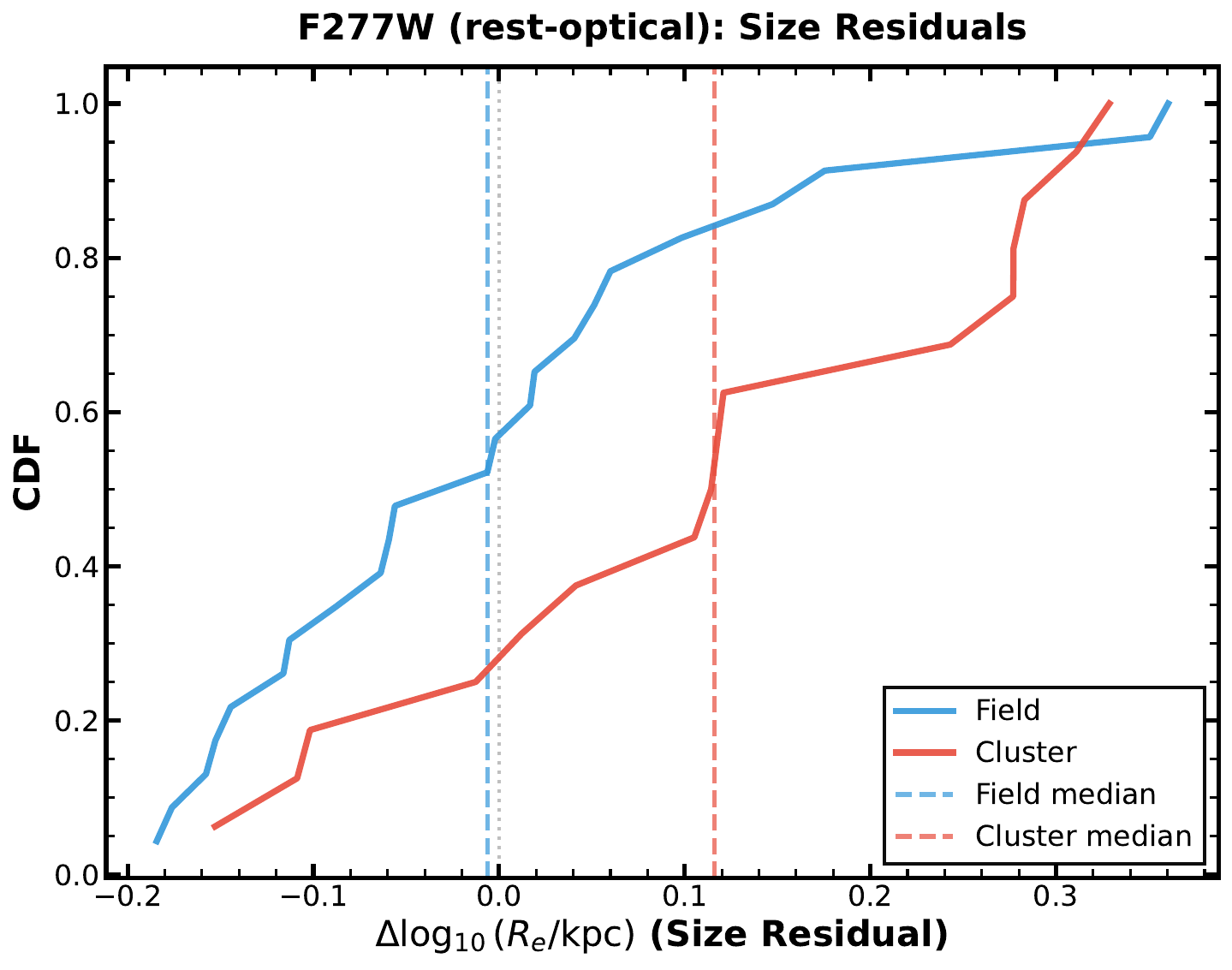}

\caption{Cumulative distribution functions (CDFs) of size residuals $\Delta\log_{10}(R_e)$ from the field size-mass relation for rest-frame UV (F150W, left) and rest-frame optical (F277W, right). Residuals are calculated as the observed size minus the size predicted from the linear relation fit to our field LAE sample (Section~\ref{sec:results_sizemass}) at each galaxy's stellar mass. Blue lines show field LAEs (centered near zero by construction), and red lines show protocluster LAEs. In rest-UV (left panel), the distributions are statistically indistinguishable (KS $p = 0.79$; Mann-Whitney $p = 0.42$). In rest-optical (right panel), protocluster LAEs show positive residuals (median $\Delta\log_{10}(R_e) = +0.12_{-0.04}^{+0.07}$~dex; Mann-Whitney $p = 0.033$, $2.1\sigma$; KS $p = 0.029$), indicating they are $\simeq$31\% larger than field galaxies of comparable stellar mass.}

\label{fig:size_cdf}

\end{figure*}

\begin{figure}
\includegraphics[width=0.99\columnwidth]{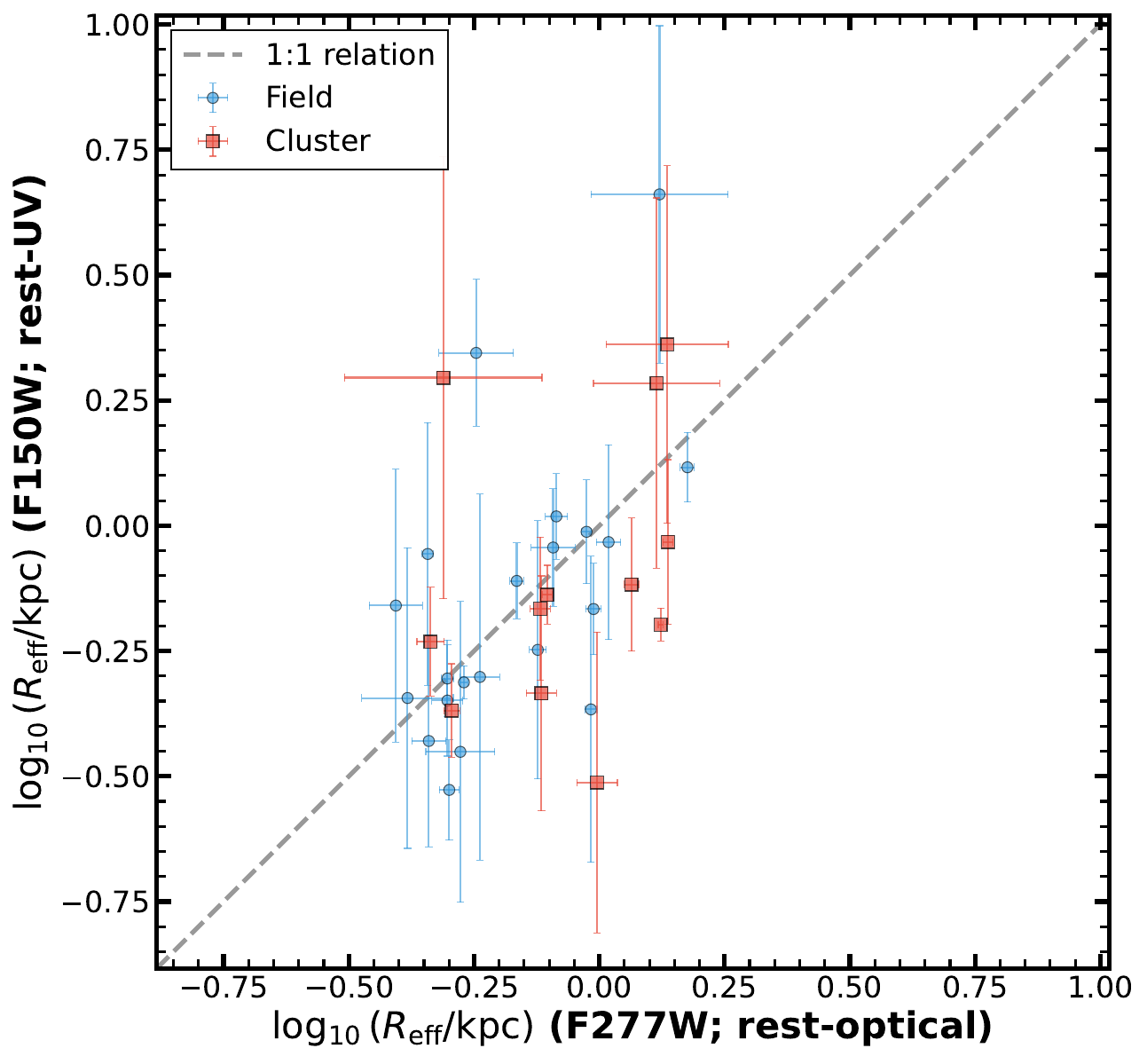}

\caption{Comparison of rest-UV (F150W) and rest-optical (F277W) effective radii for protocluster (red squares) and field (blue circles) LAEs with valid size measurements in both bands. The dashed line indicates the 1:1 relation; galaxies below the line have more compact UV emission relative to their optical extent ($R_{e,{\rm UV}} < R_{e,{\rm opt}}$). Field LAEs are broadly consistent with the 1:1 relation within their scatter. Protocluster LAEs tend to lie below the 1:1 line, but this difference is not statistically significant. This is consistent with the rest-optical size enhancement identified in Figures~\ref{fig:size_mass}--\ref{fig:size_cdf}.}

\label{fig:uv_optical_size}

\end{figure}

%==============================================================================
\section{\textbf{Results}}
\label{sec:results}

\subsection{Morphological Properties} \label{sec:results_morph}

We compare the structural properties of protocluster and field LAEs using S\'ersic profile fits in both rest-frame UV (F150W) and rest-frame optical (F277W) imaging (Figures~\ref{fig:size_mass} and \ref{fig:size_cdf}). The effective radii reveal a wavelength-dependent environmental signal. Specifically, protocluster LAEs remain $\sim$19--33\% larger than field LAEs at all tested membership radii, though the statistical significance decreases at larger projected distances (Mann-Whitney $p = 0.095$ at 4.0~pMpc, $p = 0.154$ at 5.0~pMpc), with the signal strongest at 2.0~pMpc ($p = 0.011$, $2.6\sigma$), consistent with an environmental effect concentrated in the densest core of the primary overdensity peak. For the sensitivity test including unresolved sources (assigned $R_e = 0.384$~kpc, the PSF HWHM at $z = 4.90$), protocluster LAEs remain $\sim$31\% larger than field LAEs in the expanded sample, though the formal significance decreases ($p = 0.154$). The unresolved protocluster sources have systematically lower stellar masses than the resolved ones (median $\log(M_*/M_\odot) = 8.19$ vs.\ $8.50$; Mann-Whitney $p = 0.048$), consistent with the mass-size correlation whereby lower-mass galaxies tend to be more compact and therefore more likely to fall below the PSF resolution threshold. This suggests that the exclusion of unresolved sources may be driven primarily by their lower stellar masses rather than by a systematic selection against large protocluster members, though we cannot fully exclude the possibility that the resolved and unresolved subsamples differ in other properties not examined here.

At rest-optical wavelengths, the PSF-resolved members of the primary protocluster peak show a tendency toward larger effective radii, with median $R_e = 0.81_{-0.04}^{+0.26}$~kpc compared to $R_e = 0.58_{-0.04}^{+0.11}$~kpc for field LAEs, corresponding to a $\simeq$40\% increase in median size. A Mann-Whitney $U$ test yields $p = 0.041$ ($2.0\sigma$), suggesting a difference at marginal significance. LAEs associated with the three secondary overdensity peaks show rest-optical sizes comparable to or slightly above the field median, but the differences are not statistically significant for any individual peak (Mann-Whitney $p = 0.31$--$0.79$), in contrast to the size enhancement concentrated in the primary protocluster members.

In rest-UV, this size difference is absent. Protocluster LAEs have median effective radius $R_e = 0.68_{-0.10}^{+0.08}$~kpc, compared to $R_e = 0.53_{-0.06}^{+0.20}$~kpc for field LAEs. A Mann-Whitney $U$ test yields $p = 0.51$ ($0.7\sigma$), indicating no statistically significant difference in rest-UV sizes between protocluster and field environments. This wavelength-dependent result indicates that protocluster LAEs possess more extended stellar distributions at rest-optical wavelengths, while no significant environmental difference is detected in rest-UV sizes tracing recent star formation.

In contrast to the effective radii, S\'ersic indices show no statistically significant environmental dependence in either band. In rest-optical, protocluster and field LAEs have comparable median values of $n = 0.57_{-0.25}^{+0.15}$ and $n = 0.65_{-0.20}^{+0.08}$, respectively (Mann-Whitney $p = 0.70$); in rest-UV, the corresponding values are $n = 0.52_{-0.22}^{+0.79}$ and $n = 0.43_{-0.02}^{+0.24}$ ($p = 0.40$). This indicates that the central concentration of LAE light profiles is not measurably affected by the protocluster environment at this epoch. We do not detect significant environmental differences in Ly$\alpha$ equivalent width ($p = 0.42$), UV luminosity ($p = 0.24$), or specific star formation rate ($p = 0.48$). We confirmed that using F115W (rest-NUV $\sim$1950\AA) instead of F150W yields a consistent result: protocluster LAEs show slightly larger median sizes but the difference is not statistically significant ($p = 0.12$).

\subsection{Size-Mass Relation and Residual Analysis} \label{sec:results_sizemass}

To account for the well-established correlation between galaxy size and stellar mass, we examined the size-mass relations separately for each environment (Figure~\ref{fig:size_mass}). For the field sample, we fit a linear relation of the form $\log_{10}(R_e/{\rm kpc}) = \alpha + \beta \log_{10}(M_*/M_\odot)$ to the F277W measurements using ordinary least squares. The field relation yields a slope $\beta = 0.12 \pm 0.06$ and intercept $\alpha = -1.2 \pm 0.5$ with a scatter of $0.15$ dex. Binned median sizes in equal-width stellar mass bins are shown in Figure~\ref{fig:size_mass} as large filled symbols (circles for field, squares for protocluster LAEs), with error bars representing bootstrap uncertainties and horizontal bars indicating bin widths.

Figure~\ref{fig:size_mass} reveals that protocluster LAEs systematically lie above the field size-mass relation at rest-frame optical wavelengths. For comparison, we include rest-optical size-mass relations from \citet{Song_et_al_2026} (LAEs, $z \simeq 4$--7) and \citet{Allen_et_al_2025} (SFGs, $z \simeq 4$--6); our field LAEs are broadly consistent with the former, while our protocluster LAEs are shifted above it, indicating rest-optical sizes enhanced beyond the typical LAE size-mass scaling. In rest-frame UV (F150W), the field relation has $\beta = 0.05 \pm 0.08$ and $\alpha = -0.6 \pm 0.7$ with a scatter of 0.2~dex. For context, the left panel of Figure~\ref{fig:size_mass} includes rest-UV size-mass reference lines from \citet{Allen_et_al_2025} for SFGs at $4 \leq z < 5$ and from \citet{Morishita_et_al_2024} for galaxies at $5 < z < 14$; our field LAE UV sizes are broadly consistent with these relations within the scatter. In F150W, no systematic offset is detected between protocluster and field LAEs, although the larger intrinsic scatter in rest-UV ($0.2$~dex, consistent with \citealt{Allen_et_al_2025}) limits the sensitivity to environmental differences at these wavelengths.

To quantify deviations from the expected size-mass scaling, we computed size residuals for each galaxy as $\Delta\log_{10}(R_e) = \log_{10}(R_{e,{\rm obs}}) - \log_{10}(R_{e,{\rm pred}})$, where $R_{e,{\rm pred}}$ is the size predicted from the linear relation fit to our field LAE sample in each respective filter (Section~\ref{sec:results_sizemass}) at the galaxy's stellar mass. Figure~\ref{fig:size_cdf} presents the cumulative distribution functions (CDFs) of size residuals for both samples. In F277W, protocluster LAEs exhibit predominantly positive residuals (median $\Delta\log_{10}(R_e) = +0.12_{-0.04}^{+0.07}$~dex), indicating they are systematically $\simeq$31\% larger than field galaxies of comparable stellar mass. The field sample is centered near zero by construction (median $\Delta\log_{10}(R_e) = -0.01_{-0.05}^{+0.02}$~dex). The difference between the two median residuals is $+0.12$~dex (68\% CI: $[+0.08, +0.21]$). Both a Mann-Whitney $U$ test ($p = 0.033$, $2.1\sigma$) and a Kolmogorov-Smirnov test ($p = 0.029$) indicate that the residual distributions differ, with 75\% of protocluster LAEs showing positive residuals compared to 44\% of field LAEs.

In F150W, the residual distributions for protocluster and field samples are statistically indistinguishable (Mann-Whitney $p = 0.42$, $0.8\sigma$), with protocluster median residual $+0.05$~dex and field median $-0.04$~dex. This is consistent with no size difference at rest-UV wavelengths after accounting for stellar mass. The wavelength-dependent behavior indicates that the environmental size enhancement is driven by processes affecting extended stellar populations traced by rest-optical light, rather than the rest-UV sizes tracing recent star formation, where no significant environmental difference is detected.
A direct comparison for galaxies with reliable sizes in both bands (13 protocluster and 21 field LAEs passing PSF cuts in both F150W and F277W; Figure~\ref{fig:uv_optical_size}) is consistent with this result: protocluster LAEs tend to lie below the 1:1 relation, but the difference in UV-to-optical size ratio between environments is not statistically significant.

\section{\textbf{Discussion}}
\label{sec:discussion}

\subsection{Comparison with LAE Size Measurements at $z \simeq 5$} \label{sec:disc_lae_morph}

Our field LAE sample at $z = 4.90$ shows median rest-optical (F277W) effective radius $R_e = 0.58_{-0.04}^{+0.11}$~kpc, consistent with the characteristically compact sizes ($R_e \lesssim 1$~kpc) of LAEs observed across a wide range of redshifts \citep{Kim_et_al_2026}. Stacking analysis of LAEs at $z \approx 5.7$ in the COSMOS field revealed disk-like light profiles (S\'ersic $n \sim 0.7$; \citealt{Taniguchi_et_al_2009}), consistent with the low S\'ersic indices found in our sample. This is also in agreement with the size-mass relation measured by \citet{Song_et_al_2026} for LAEs at $4 < z < 7$, who found that LAE sizes closely match those of typical SFGs in both slope and normalization at these redshifts, with statistically comparable UV and optical sizes ($\log(R_{e,V}/R_{e,{\rm UV}}) \approx 0.03$).

In contrast, our protocluster LAEs exhibit a larger median $R_e = 0.81_{-0.04}^{+0.26}$~kpc, a $\sim$40\% increase over the field. This size enhancement occurs without a corresponding change in S\'ersic index ($p = 0.70$), indicating that protocluster LAEs maintain similar light profile shapes to field LAEs while possessing more extended stellar distributions in the rest-optical.

\subsection{Environmental Effects on Galaxy Morphology in Protoclusters} \label{sec:disc_environment}

While environmental effects on galaxy morphology have been studied in several protoclusters at $z \lesssim 3$, no previous work has examined the morphological properties of LAEs in rest-frame optical in protocluster environments at $z \gtrsim 4$. At lower redshifts, rest-optical HST WFC3 F160W observations of 151 H$\alpha$ emitters (HAEs) in protoclusters at $z \simeq 2.23$ found similar S\'ersic index distributions between protocluster and field populations \citep{GoldenMarx_et_al_2025}, consistent with the absence of environmental trends in rest-UV morphology for UV-selected star-forming galaxies at similar redshifts \citep{Peter_et_al_2007}. The morphology-density relation was found to be already in place at $z \simeq 2$ in CARLA clusters at $1.4 < z < 2.8$ \citep{Mei_et_al_2023}. At $z \simeq 1.5$, [O~{\sc ii}] emitters in an extreme overdensity showed a higher frequency of morphological disturbances and relatively larger sizes in the densest regions compared to the field, along with elevated star formation activity \citep{Laishram_et_al_2024}. At $z = 4.3$, ALMA dust continuum observations of dusty star-forming galaxies (DSFGs) in the SPT2349$-$56 protocluster core found no significant difference in half-light radii compared to field DSFGs \citep{Hill_et_al_2020}. Image stacking of proto-brightest cluster galaxy (proto-BCG) candidates at $z \simeq 4$ revealed $\sim$28\% larger rest-UV effective radii compared to magnitude-matched field galaxies \citep{Ito_et_al_2019}, although this enhancement applies to the most luminous UV-selected protocluster members, a population distinct from the lower-mass LAEs studied here. \citet{Overzier_et_al_2008} studied LAEs in the TN~J1338$-$1942 protocluster at $z = 4.1$ using \textit{HST} ACS rest-UV imaging and found no evidence for environmental differences in sizes or morphologies, although the broad photometric redshift selection ($\Delta z \sim 0.5$) may dilute any environmental signal. This rest-UV null result is consistent with our finding of no significant size difference at rest-UV wavelengths ($p = 0.51$), suggesting that the environmental signature we detect is accessible only at rest-frame optical wavelengths. Our study extends these efforts to $z = 4.90$ with the addition of JWST rest-frame optical imaging, revealing that the environmental size enhancement in protocluster LAEs emerges specifically in the extended stellar populations traced by longer wavelengths --- a signal that was inaccessible to previous rest-UV-only studies.

Environmental dependence of galaxy sizes has been established at lower redshifts: a study of $>3$ million galaxies (stellar mass $\log M_*/M_\odot \gtrsim 8.9$--$9.9$, with the lower completeness limit increasing with redshift) from Hyper Suprime-Cam at $0.3 < z < 0.7$ found galaxies in denser environments are $\sim$10--20\% larger in rest-optical than counterparts matched in stellar mass and morphology in lower-density regions \citep{Ghosh_et_al_2024}. However, these studies focused on massive, mass-selected galaxy populations at $z \lesssim 1$. Our observed $\sim$40\% rest-optical size enhancement at $z = 4.90$ ($\sim$31\% at fixed stellar mass from residual analysis) represents the first such measurement for LAEs in a protocluster environment. Direct comparison with the low-redshift results is limited by differences in galaxy populations (LAEs vs.\ mass-selected samples) and selection methods, but the detection of an environmental size signal at this early epoch suggests that processes influencing galaxy structure are already operating in protocluster environments during the first $\sim$1.2~Gyr of cosmic history.
 
\subsection{Size-Mass Relations in Different Environments} \label{sec:disc_sizemass}

Recent JWST studies have established the rest-optical size-mass relation for high-redshift galaxies, with slopes consistent across different populations and redshifts \citep{Song_et_al_2026, Allen_et_al_2025}. Our field LAE relation is consistent with these results within the uncertainties (Section~\ref{sec:results_sizemass}). The key environmental result is in the normalization: at fixed stellar mass, protocluster LAEs are systematically offset toward larger rest-optical sizes.

At lower redshifts, some studies of star-forming galaxies in protocluster environments have found no significant environmental effect on the size-mass relation: H$\alpha$ emitters (HAEs) in protoclusters at $z \simeq 2$ show similar rest-frame UV sizes to field HAEs at fixed stellar mass \citep{Naufal_et_al_2023}, and mass-matched star-forming samples at $z < 0.5$ show no rest-optical size difference between cluster and field environments \citep{Chen_et_al_2024}. Our results at $z = 4.90$ show that even among star-forming LAEs, the protocluster environment produces a measurable offset in rest-optical sizes at fixed stellar mass (Figure~\ref{fig:size_mass}), with a median residual of $\Delta\log_{10}(R_e) = +0.12$~dex (68\% CI: $[+0.08, +0.21]$). This suggests that environmental size enhancement may already be in place for star-forming populations at this epoch, preceding the quiescent-dominated effects seen at $z < 2$.

\subsection{Wavelength-Dependent Size Measurements} \label{sec:disc_wavelength}

\citet{Allen_et_al_2025} measured rest-optical and rest-UV size-mass relations for star-forming galaxies at $z = 3\mbox{--}9$ and found that the rest-UV size-mass relation has both a steeper slope and larger intrinsic scatter ($\sigma \sim 0.22\mbox{--}0.25$~dex) compared to rest-optical ($\sigma \sim 0.17\mbox{--}0.19$~dex), with low-mass galaxies at $z > 5$ being $>2\sigma$ smaller in rest-UV than in rest-optical. In contrast, \citet{Ono_et_al_2024} found the average $r_{e,{\rm opt}}/r_{e,{\rm UV}} \approx 1.01$ across $z = 4\mbox{--}10$, suggesting that on average UV and optical sizes remain comparable at these epochs.
Our result shows a wavelength-dependent environmental signature: protocluster LAEs are significantly larger in rest-optical but not in rest-UV compared to field LAEs. This contrasts with the general LAE population at $z \gtrsim 4.9$, where rest-UV and rest-optical sizes are statistically comparable \citep{Song_et_al_2026}, suggesting that the protocluster environment introduces a wavelength-dependent structural difference not present in the general field population.

The physical basis for wavelength-dependent sizes reflects differences in stellar populations: rest-UV light traces young, massive stars formed over the last $\sim$100 Myr, while rest-optical light is more sensitive to older stellar populations \citep{Papovich_et_al_2005}. Studies of stellar population gradients have shown these are the dominant factor ($\sim$80\%) driving wavelength-dependent effective radii, with differential dust attenuation contributing secondarily \citep{Baes_et_al_2024}. In this context, the wavelength-dependent environmental signature we observe (Section~\ref{sec:results_morph}) suggests that protocluster processes preferentially affect the extended older stellar populations rather than the central regions where current star formation is concentrated. Tidal interactions among neighboring galaxies can redistribute existing stellar material to larger radii, while accretion of smaller companions \citep[e.g.,][]{Shimizu_Umemura_2010} may deposit stars preferentially in the galaxy outskirts, building up an extended rest-optical envelope around the compact star-forming core. Additionally, an earlier onset of star formation in the overdense environment could allow more time to accumulate spatially extended older stellar populations traced by rest-optical light. To examine whether an age offset accompanies the size difference, we compared stellar ages derived from SED fitting between protocluster and field LAEs and find no statistically significant difference in the stellar age distributions between the two populations. We caution, however, that stellar age estimates from broadband SED fitting at $z \sim 5$ carry substantial uncertainties due to the age--metallicity degeneracy and outshining by young stellar populations; the present data are therefore insufficient to conclusively constrain the earlier-onset scenario. With the present data, we cannot distinguish among these scenarios, and multiple processes may operate simultaneously. Disentangling their relative contributions will require larger LAE samples across multiple protoclusters, complemented by spectroscopic observations at these redshifts. The absence of a significant environmental difference in rest-UV sizes is consistent with the characteristically compact UV morphologies of LAEs observed across $z \simeq 2$--$6$ \citep{PaulinoAfonso_et_al_2018, Shibuya_et_al_2019, Song_et_al_2026}. Indeed, compact rest-UV morphology is closely linked to Ly$\alpha$ photon escape, as LAEs with higher equivalent widths tend to be smaller and more concentrated \citep{PaulinoAfonso_et_al_2018}; this implies that this rest-UV null result may partly reflect a selection effect, since LAEs are by definition drawn from the compact end of the UV size distribution. Moreover, Ly$\alpha$ photons in evolved LAEs can escape through low-density channels in the ISM cleared by outflows \citep{Shimizu_et_al_2025}; in the protocluster environment, tidal interactions may similarly disrupt the ISM, potentially facilitating Ly$\alpha$ escape while also producing the larger rest-optical sizes we observe.

Future wide-field spectroscopic surveys (e.g., Subaru/PFS) combined with JWST imaging of additional protoclusters at similar redshifts will be needed to determine whether this wavelength-dependent size enhancement is a general feature of LAEs in overdense environments. Spatially resolved spectroscopy (e.g., JWST/NIRSpec IFU) would further enable measurements of stellar population gradients and gas kinematics to constrain the physical processes driving the observed size differences.

\section{\textbf{Conclusions}}
\label{sec:conclusions}

We report the discovery of the Loktak protocluster at $z = 4.90$ in the COSMOS field and present one of the first systematic rest-frame optical morphological comparisons between protocluster and field LAEs at this epoch, using JWST NIRCam imaging from COSMOS-Web. Our main findings are:

(1) We identify a large-scale overdensity structure spanning $\sim$65 $\times$ 36 cMpc$^2$, comprising four distinct peaks with surface density enhancements of up to $\sim$4 times the field density within 1.5~pMpc radii, with the primary peak enclosing 18 LAEs within this radius.

(2) The PSF-resolved members of the primary protocluster peak show a tendency toward larger effective radii in rest-frame optical wavelengths (median $R_e = 0.81$~kpc vs.\ $0.58$~kpc; Mann-Whitney $p = 0.041$, $\sim$40\% larger; when unresolved sources are included with an upper-bound size of 0.384~kpc, the cluster remains $\sim$31\% larger, though formal significance decreases ($p = 0.154$)). No significant size difference is found in rest-frame UV ($p = 0.51$). The enhancement is strongest at 2.0~pMpc ($p = 0.011$, $2.6\sigma$) and weakens at larger radii, consistent with a density-driven, core-concentrated effect.

(3) The size-mass residual analysis supports this result: at fixed stellar mass, protocluster LAEs in F277W are offset by $\Delta\log_{10}(R_e) = +0.12$~dex (68\% CI: $[+0.08, +0.21]$; Mann-Whitney $p = 0.033$, $2.1\sigma$; KS $p = 0.029$) above the field relation. S\'ersic indices show no significant environmental dependence ($p = 0.70$ in F277W).

(4) This wavelength-dependent size enhancement suggests that protocluster environments at $z \simeq 5$ preferentially influence extended stellar populations traced by rest-optical light, with no significant environmental difference detected in the rest-UV sizes tracing recent star formation.

Extending these measurements to additional protoclusters across $4 < z < 6$, with larger samples and deeper imaging, will be essential to determine whether this environmental signature is a universal feature of LAE evolution in overdense regions or specific to this system.

\section{\textbf{Data Availability}}

The JWST COSMOS-Web imaging and photometric catalog used in this work were retrieved from the COSMOS2025 data release, available at \url{https://cosmos2025.iap.fr}. The NB718 LAE catalog from the SILVERRUSH survey \citep{Kikuta_et_al_2023} will be made public following the HSC-SSP Public Data Release 4 (PDR4).

\begin{acknowledgments}
We thank the anonymous referee for constructive comments that improved the clarity and scientific content of this manuscript. This work was supported by JSPS KAKENHI grant No. J23H01219 and JSPS Core-to-Core Program (grant No.: JPJSCCA20210003). H.K. acknowledges support from JSPS KAKENHI grant Nos. 23KJ2148 and 25K17444. SK acknowledges support from JSPS KAKENHI grant Nos. 24KJ0058 and 24K17101. TK and YK acknowledge financial support from JSPS KAKENHI Grant Numbers 24H00002 (Specially Promoted Research by T. Kodama et al.), 22K21349 (International Leading Research by S. Miyazaki et al.).

This research is based in part on data collected at the Subaru Telescope, which is operated by the National Astronomical Observatory of Japan. We are honored and grateful for the opportunity of observing the Universe from Maunakea, which has the cultural, historical, and natural significance in Hawaii.

The Hyper Suprime-Cam (HSC) collaboration includes the astronomical communities of Japan and Taiwan, and Princeton University. The HSC instrumentation and software were developed by the National Astronomical Observatory of Japan (NAOJ), the Kavli Institute for the Physics and Mathematics of the Universe (Kavli IPMU), the University of Tokyo, the High Energy Accelerator Research Organization (KEK), the Academia Sinica Institute for Astronomy and Astrophysics in Taiwan (ASIAA), and Princeton University. Funding was contributed by the FIRST program from the Japanese Cabinet Office, the Ministry of Education, Culture, Sports, Science and Technology (MEXT), the Japan Society for the Promotion of Science (JSPS), Japan Science and Technology Agency (JST), the Toray Science Foundation, NAOJ, Kavli IPMU, KEK, ASIAA, and Princeton University.
The HSC data were retrieved from the HSC data archive system, which is operated by Subaru Telescope and Astronomy Data Center (ADC) at NAOJ.
\end{acknowledgments}

\bibliography{sample631}{}
\bibliographystyle{aasjournal}

\end{document}